\documentclass[prd,superscriptaddress]{revtex4}
\input epsf
\usepackage{graphicx}
\usepackage[centertags]{amsmath}
\usepackage{amsfonts}
\usepackage{amssymb}
\usepackage{amsthm}
\usepackage{newlfont}
\usepackage{float}
\usepackage{longtable}
\usepackage[english]{babel}

\newcommand{\HH}{{\cal H}}
\begin{document}
\selectlanguage{english}
\title{Matter density perturbations in interacting quintessence models}
\author{Germ\'{an} Olivares\footnote{E-mail address: german.olivares@uab.es}}
\affiliation{Departamento de F\'{\i}sica, Universidad Aut\'{o}noma de Barcelona,
Barcelona, Spain}
\author{Fernando Atrio-Barandela\footnote{E-mail address: atrio@usal.es; on
sabbatical leave from Departamento de F\'{\i}sica Fundamental,
Universidad de Salamanca, Spain}}
\affiliation{Department of Physics and Astronomy, University of Penssylvania, Philadelphia,
PA 19104}
\author{Diego Pav\'{o}n\footnote{E-mail address: diego.pavon@uab.es}}
\affiliation{Departamento de F\'{\i}sica, Universidad Aut\'{o}noma de Barcelona,
Barcelona, Spain}

\begin{abstract}
Models with dark energy decaying into dark matter have been
proposed to solve the coincidence problem in cosmology. We study
the effect of such coupling in the matter power spectrum. Due to
the interaction, the growth of matter density perturbations during
the radiation dominated regime is slower compared to
non-interacting models with the same ratio of dark matter to dark
energy today. This effect introduces a damping on the power
spectrum at small scales proportional to the strength of the
interaction, $c^2$, and similar to the effect generated by
ultrarelativistic neutrinos. The interaction also shifts
matter--radiation equality to larger scales. We compare the matter
power spectrum of interacting quintessence models with the
measurments of 2dFGRS. The data are insensitive to values of
$c^2\le 10^{-3}$ but strongly constraints larger values. We
particularize our study to models that during radiation domination
have a constant dark matter to dark energy ratio.
\end{abstract}
\maketitle

\section{Introduction}
Observations of high redshift supenovae \cite{riess}, temperature
anisotropies of the cosmic background radiation \cite{wmap,wmap3},
matter power spectrum \cite{sdss,2dgf}, and the integrated Sachs--Wolfe
signal \cite{isw} indicate that the Universe is currently
undergoing a phase of accelerated expansion \cite{reviews}. A
cosmological constant, $\Lambda$, is frequently invoked as the
most natural candidate to drive this acceleration. However, this
choice  is rather problematic. First, the observed $\Lambda$ value
falls by many orders of magnitude below the prediction of quantum
field theories \cite{weinberg}. Second, it is hard to understand
why precisely today the vacuum energy density is of the same order
of magnitude than that of matter. This remarkable fact, known as
the ``coincidence problem'' \cite{coincidence}, lacks of a fully
convincing theoretical explanation.

Models based on at least two matter components (baryonic and dark)
and one  dark energy component (with a high negative pressure)
have been suggested  to explain the accelerated rate of expansion
and simultaneously alleviate the coincidence problem
\cite{peebles_rmph,dualk}. Further, coupling between dark matter
(DM) and  dark energy (DE) has been suggested as a possible
explanation for the coincidence problem \cite{amendola,rong}. In
particular, the interacting quintessence models of references
\cite{amendola,dualk}, require the ratio of matter and dark energy
densities to be constant at late times. The coupling between
matter and quintessence is either motivated by high energy
particle physics considerations \cite{amendola} or is constructed
by requiring the final matter to dark energy ratio to be stable
against perturbations \cite{iqm0,iqm,domainw}. The nature of both
DM and DE being unknown, there is no physical argument to exclude
their interaction. On the contrary, arguments in favor of such
interaction have been suggested \cite{peebles}, and more recently
they have been extended to include neutrinos \cite{tocchini}. As a
result of the interaction, the matter density drops with the scale
factor $a(t)$ of the Friedmann--Robertson--Walker metric more
slowly than $a^{-3}$. The interacting quintessence model studied
in the literature have been found to agree with observations of
WMAP data and supernovae \cite{olivares}, but they require values
of cosmological parameters different from those of WMAP (first
year) concordance model. Observations of the large scale structure
can also be used to constrain models. Recent data includes the
SDSS \cite{sdss} and 2dFGRS \cite{2dgf} measurements of the matter
power spectrum. The analysis of 2dFGRS showed discrepancies with
WMAP first year data but is in much closer agreement than SDSS
with the results of WMAP third year data \cite{wmap}.

Currently, there is no compelling evidence for DM--DE interaction
\cite{szydlowski} and its (non-)existence must be decided
observationally.  It has been suggested that the skewness of the
large scale matter distribution is a sensitive parameter to
determine the difference in the clustering of baryons and dark
matter resulting from the interaction \cite{amendola_prl}. In this
paper we shall study the effect of the interaction on the
evolution of matter density perturbations during the radiation
dominated period. The evolution of matter and radiation density
perturbations provides powerful tools to constrain the physics of
the dark sector \cite{koivisto}. We will show how the shape of the
matter power spectrum can be a directly related with the
interaction and we will use the matter power spectrum measured by
the 2-degree field galaxy redshift survey (2dFGRS, for short) to
set constraints on the interaction. The outline of the paper is as
follows: In Section II, we present a brief summary of the
interacting quintessence model (IQM, hereafter). In Section III we
describe the evolution of matter and radiation perturbations in
models with dark matter and dark energy. In Section IV we discuss
some analytical solutions and in Section V we show how the slope of
a scale invariant matter density perturbations has less power on small scales than
non-interacting models. In Section VI we describe the results
of Monte Carlo Markov chains constructed to compare the model with
the observations. Finally, Section VII summarizes our main results
and conclusions.

\section{The interacting quintessence model \label{sec2}}
Most cosmological models implicitly assume that matter and dark
energy interact only gravitationally. In the absence of an
underlying symmetry that would suppress a matter -- dark energy
coupling (or interaction) there is no a priori reason for
dismissing it. Cosmological models in which dark energy and matter
do not evolve separately but interact with one another were first
introduced to justify the small value of the cosmological constant
\cite{wetterich}. Recently, various proposals at the fundamental
level, including field Lagrangians, have been advanced to account
for the coupling \cite{federico}. Scalar field Lagrangians coupled
to matter generically do not generate scaling solutions with a
long enough dark matter dominated period as required by structure
formation \cite{amendola-challenges}. The phenomenological model
we will be considering was constructed to account for late
acceleration in the framework of Einstein relativity and to
significantly alleviate the aforesaid coincidence problem
\cite{iqm0,iqm} and escapes the limits imposed by
\cite{amendola-challenges}. Here we shall describe its main
features. For further details see  Refs.
\cite{iqm,dualk,section2}.

The model considers a spatially flat Friedmann--Robertson--Walker
universe filled with radiation, baryons, dark matter (subscript,
$c$) and dark energy (subscript, $x$). Its key assumption is that
dark matter and dark energy are coupled by a term $Q = 3 H c^{2}
(\rho_{c} +\rho_{x})$, that gauges the transfer of energy from the
dark energy $\rho_x$ to the dark matter $\rho_c$. The quantity
$c^{2}$ is a small dimensionless constant that measures the
strength of the interaction, and $ H= a^{-1}\, da/dt$ is the
Hubble function. To satisfy the severe constraints imposed by
local gravity experiments \cite{peebles_rmph, hagiwara}, baryons
and radiation couple to the other energy components only through
gravity. Thus, the energy balance equations for dark matter and
dark energy take the form
\begin{equation}
\frac{d \rho_{c}}{dt}+3 H \rho_{c} = Q\,  ,\qquad
\frac{d\rho_{x}}{dt}+3 H (1+w_{x})\rho_x= -Q \, ,
\label{balance}
\end{equation}
where $w_{x} = p_{x}/\rho_{x} < -1/3$ is the equation of state
parameter of the dark energy fluid.

Our ansatz for $Q$ guarantees that the ratio between energy
densities $r \equiv \rho_{c}/\rho_{x}$ tends to a fixed value at
late times. This can be seen by studying the evolution of $r$
which is governed by
\begin{equation}
\frac{dr}{dt} = - 3 \Gamma H r \, , \quad \qquad \Gamma = -w_{x}- c^{2}%
\frac{(\rho_{c}+\rho_{x})^{2}}{\rho_{c}\, \rho_{x}} \, .
\label{r-evolution}
\end{equation}
The stationary solutions of Eq. (\ref{r-evolution}) follow from
solving $r_{s} \, \Gamma(r_{s}) = 0$. When $w_{x}$ is a constant
these solutions are given by the roots of the quadratic expression
\begin{equation}
r^{\pm}_{s} = -1 + 2b \pm 2 \sqrt{b(b-1)}\, ,  \qquad b = -%
\frac{w_{x}}{4c^{2}} > 1.
\label{r+-}
\end{equation}
As it can be checked by slightly perturbing Eq.
(\ref{r-evolution}), the stationary $r^{+}_{s}$ solution is
unstable while $r^{-}_{s}$ is stable. The general solution of Eq.
(\ref{r-evolution}) can be written as
\begin{equation}
r(x) = \frac{r^{-}_{s}+ x r^{+}_{s}}{1 + x}\, ,
 \label{r(x)}
\end{equation}
where $x = (a/a_{0})^{-\lambda}$ with $\lambda \equiv%
12\, c^{2}\sqrt{b(b-1)}$. In the range $r^{-}_{s} < r < r^{+}_{s}$
$r(x)$ is  a monotonic decreasing function. Thus, as the Universe
expands, $r(x)$ gently evolves from $r_{s}^{+}$ to the attractor
solution $r^{-}_{s}$. The transition from one asymptotic solution
to the other occurs only recently (see Fig. 2 of  \cite{olivares})
so we can take $r\simeq r^{+}_{s}$ during a fairly large part of
the history of the Universe. Finally, the constraint on $r_{s}^{+}
\simeq$ const implies that $r$ and $c^2$ are not independent, but
linked by $c^2(r_{s}^{+}\, +1)^2= r_{s}^{+}\, \mid w_{x}\mid$, so
the product $c^{2}r_{s}^{+} \sim \mid w_{x}\mid$ is of order
unity.

We would like to remark that the above ansatz for $Q$ is not
arbitrary. It was chosen so that the ratio between dark matter and dark
energy densities tends to a fixed value at late times, thereby
alleviating the coincidence problem \cite{iqm,dualk,section2}. It
also yields a constant but unstable ratio at
early times. It is hard to imagine a simpler expression for $Q$
entailing these two key properties. Likewise, it is only fair to
acknowledge that the aforesaid expression can be re--interpreted
as implying, at late times, an effective exponential potential for
the quintessence field. This well--known result was derived by
Zimdahl {\em et al.} \cite{iqm0}. Likewise, in \cite{olivares} we
remarked that the effective potential of the IQM model exhibits a
power--law dependence on the quintessence field at early times and
an exponential dependence at late times.

Near  $r \approx r^{-}_{s}$ the balance Eqs. (\ref{balance}) can
be approximated by
\begin{eqnarray}
\frac{1}{\rho_{c}}\, \frac{d \rho_{c}}{d r} &\simeq & \frac{1-c^{2} (1+%
1/r^{-}_{s})}{c^{2}(r^{+}_{s} - r^{-}_{s}) (r -r^{-}_{s} )} \, ,\nonumber \\
\frac{1}{\rho_{x}}\, \frac{d \rho_{x}}{d r} & \simeq &%
\frac{1+w_{x}+c^{2} (1+r^{-}_{s}}{c^{2}(r^{+}_{s} - r^{-}_{s}) (r%
-r^{-}_{s} )} \, \, .
\label{near-attractor}
\end{eqnarray}
For $w_{x}\simeq$ constant, these equations can be integrated to
\\
\begin{equation}
\rho_{c} \propto a^{-3\left[ 1-c^{2}(1+ 1/r^{-}_{s})\right]}\, ,
\qquad \rho_{x} \propto a^{-3\left[1+w_{x}+c^{2}(1+r^{-}_{s})%
\right]} \, .
\label{integrate}
\end{equation}
Notice that the condition $\Gamma(r^{-}_{s}) = 0$ implies that the
exponents in the energy densities, Eq. (\ref{integrate}),
coincide.  Interestingly, these results are not only valid when
the dark energy is a quintessence field (i.e., $-1 < w_{x} <-1/3
$), they also apply when the dark energy is of phantom type (i.e.,
$w_{x} <-1$), either a scalar field with the ``wrong sign" for the
kinetic energy term, a {\em k}-essence field, or a tachyon field
\cite{dualk}.

Near the attractor, dark matter and dark energy dominate
the expansion and Friedmann equation becomes simply $3 \,H^{2} = \kappa%
(\rho_{c} + \rho_{x})$ and $a \propto%
t^{(2/3)[1+w_{x}+c^{2}(1+r^{-}_{s})]^{-1}}$. The results presented
here significantly alleviates the coincidence problem but they  do
not solve it in full. For this purpose, one needs to show that the
attractor was reached only recently -or that we are very close to
it- and that $r^{-}_{s}$ is of order unity. In fact, the value of
$r^{-}_{s}$ cannot be derived from data and must be understood as
an input parameter. This is also the case of a handful of key
cosmic quantities such as the current value of the cosmic
background temperature, the Hubble constant or the ratio between
the number of baryons and photons.

\section{Linear perturbations}
As the scalar field is coupled just to dark matter and  since dark
matter and quintessence are coupled to baryons and photons only
gravitationally, there is no transfer of energy or momentum from
the scalar field to baryons or radiation and their evolution is
the same as in non-interacting models. In the synchronous gauge
and for a flat space-time, the line element is given by: $ds^2 =
a^2(\tau)[-d\tau^2+ (\delta_{K,ij}+h_{ij})dx^idx^j]$, where $\tau$
is the conformal time, $a$ the scale factor and $\delta_{K,ij}$ is
Kronecker's delta tensor. Only two functions $h$ and $\mu$ are
necessary to characterize the scalar mode of the metric
perturbations $h_{ij}$ \cite{bertschinger}. Assuming the dark
energy energy-momentum tensor is free of anisotropic stresses, the
equations describing the dark matter and dark energy evolution in
the synchronous gauge are:
\begin{eqnarray}
\dot{\delta_x} &=& -(1+w_x)(\theta_x+\frac{\dot h}{2})-3\HH(1-w_x)\delta_x-9\HH^2(1-w_x^2)
\frac{\theta_x}{k^2} + 3{\HH}c^2(\delta_x+r\delta_c) \, ,   \label{de_cont}\\
\dot{\theta_x} &=& 2{\HH}\theta_x + \frac{k^2}{1+w_x}\delta_x
-3{\HH}\frac{c^2}{1+w_x}(1+r)\theta_x , \label{de_euler}\\
\dot{\delta_c} &=& -\theta_c-\frac{\dot h}{2}-3{\HH}c^2(\delta_c+\frac{\delta_x}{r})\, ,
\label{cdm_cont}\\
\dot{\theta_c} &=& -{\HH}\theta_c + 3{\HH}c^2(1+1/r)\theta_x .  \label{cdm_euler}
\end{eqnarray}
In these expressions, the derivatives are taken with respect to
the conformal time $\tau$, $\delta$ is the density fluctuation,
$\theta$ the divergence of the fluid velocity, $h$ the
gravitational potential, $k$ the wavenumber of a Fourier mode,
${\HH}=\dot a/a$ and, $r$ is the ratio of the background cold dark
matter to the dark energy density.  We also assume that the dark
energy has constant equation of state parameter $w_x=$ constant
and sound speed $c_{s,x}=1$. In this gauge, $(\delta
P/\delta\rho)_x\delta_x=c^2_{s,x}\delta_x + 3\HH\theta_x(1+w_{x})
(c^2_{s,x}-w_{x})/k^2$. As noted in \cite{koivisto}, Eq.
(\ref{cdm_euler}) was mistyped in \cite{olivares}.

Equations (\ref{de_cont})--(\ref{cdm_euler}) do not form a closed
set. They must be supplemented with the equations describing the
evolution of the coupled baryon--photon fluid, neutrinos and
gravitational potentials. For the potentials, the only relevant
quantity is the trace of the metric perturbation, $h$. Its time
evolution can be derived from Einstein's equations:
\begin{equation}
\ddot h+{\HH}\dot h=-3{\HH}^2\sum_{i}\left(1+3\, %
(\delta P/\delta\rho)_{s,i}^{2}\right)
\delta_i\Omega_i \, , \label{hgravitation}
\end{equation}
where the sum is over all matter fluids and scalar fields; $\Omega_i$
is the energy density of fluid $i$ in units of the critical density.
With respect to baryons, photons and neutrinos, they interact with
the DM and DE only through gravity.

\begin{figure}
\includegraphics[scale=0.9]{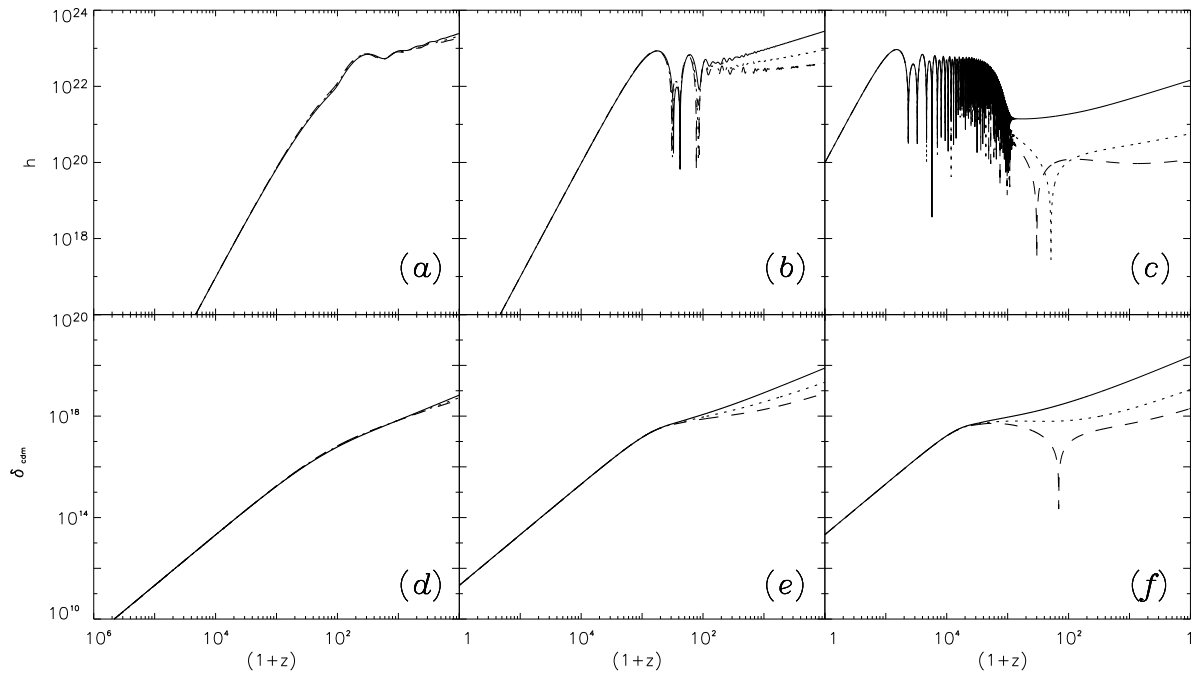}
\vspace*{-3cm} \caption{Evolution of the gravitational potential
(upper panels) and the cold dark matter density perturbations
(lower panels) for three modes: $k=0.01$ (left)
 $k=0.1$ (center) and $k=1h^{-1}$Mpc (right panels). We study the evolution
of each mode in three different cosmological models: the
concordance model ($c^2=0$, solid line) and two interacting
quintessence models with the same cosmological parameters,
$c^2=10^{-3}$ (dotted) and $c^2=6\times 10^{-3}$ (dashed
line).} \label{fig1}
\end{figure}

The coupled evolution of dark matter, dark energy, baryon, photon
and, optionally, neutrino density perturbations and gravitational
fields can not be solved analytically. To compute numerically the
solution, we have implemented Eqs.
(\ref{de_cont})--(\ref{hgravitation}) into the CMBFAST code
\cite{cmbfast}. In Fig. \ref{fig1} we show the evolution of the
potential, $h$, and the matter density perturbation, $\delta_c$,
for three modes of wavelength $k=0.01, 0.1$ and $1h$ Mpc$^{-1}$
and for three different values of the DE decay rate: $c^2=0$
(solid line), $c^2=10^{-3}$ (dotted line) and $c^2=6\times
10^{-3}$ (dashed line). In all the cases, the cosmological
parameters defining the background model are: $\Omega_c=0.26$,
$\Omega_b=0.04$, $\Omega_x=0.7$, $w_x=-0.9$ and the Hubble
constant $H_0 =70$ km/s\, Mpc$^{-1}$.

As panels \ref{fig1}e and \ref{fig1}f illustrate, modes that enter
the horizon before matter-radiation equality grow slower with
increasing interaction rate. As a result, the matter power
spectrum on those scales will have less power than in
non-interacting models.  To obtain some
insight on the behavior on the evolution of matter density
perturbations, we will be considering some limiting cases where
analytic solutions exist. For simplicity, we shall assume the
dynamical effect of baryons and neutrinos in the evolution of dark
matter and dark energy density perturbations can be neglected. The
result of combining Eqs. (\ref{cdm_cont}) and (\ref{cdm_euler}),
considering only the leading terms at first order in the
interaction $c^2$, is
\begin{equation}
\ddot \delta_c +\HH (1+3c^2)\dot\delta_c + 3c^2\HH^2
\left(\frac{1-3w_{B}}{2}+\frac{3c^2(r+1)}{(1+w_x)A}\right)\delta_c=
-{1\over 2}(\ddot h+\HH\dot h)+c^2F(\dot\delta_x,\delta_x),
\label{ho}
\end{equation}
with
\begin{eqnarray}
F(\dot \delta_x,\delta_x)&=& 3 \HH \dot\delta_x+3\frac{\HH^2}{r}
\left[\frac{3(r+1)(1-w_x-c^2)}{(1+w_x)A}-\frac{1-3w_B}{2}\right]\delta_x
\, ,
\\
A &=& 1+(3\HH/k)^2(1-w_x^2)  \, .
\end{eqnarray}
Equation (\ref{ho}) corresponds to a damped harmonic oscillator
(with real or imaginary frequencies) with a forcing term that,
since $c^2\ll 1$, is dominated by the time evolution of the
gravitational potential. In this approximation, Eq.
(\ref{hgravitation}) gives
\begin{equation}
\ddot \delta_c +\HH (1+3c^2)\dot\delta_c
-3\HH^2\left[\frac{\Omega_c}{2}\left(1+{3(1-w_x)(3\HH/k)^2\over A}c^2\right)-c^2
\left(\frac{3c^2(r+1)}{(1+w_x)A}+\frac{1-3w_B}{2}\right)
\right]\delta_c=3
\HH^2(\Omega_\gamma\delta_\gamma+2\Omega_x\delta_x)\, .
\label{ho_cdm}
\end{equation}
In the limit $c^2=0$ this equation coincides with the evolution of
matter density perturbation in non-interacting cosmologies. The
effect of the interaction is to increase the friction term and to
modify the oscillation frequency. The term in square brackets
accounts for the self attractive force acting on the perturbation;
the extra contribution arises due to the interaction.  The effect
of the DM--DE coupling can be understood as a modification of the
effective gravitational constant. This result was previously found
in \cite{amendola2004} and \cite{corasaniti}, both models with a
different interaction ansatz. In our case, the
interaction provides a new physical effect not present in other
models: if $w_x\sim -1$, the second term in the square
parenthesis could dominate and matter density perturbations would
stop growing and start oscillating.

\section{Evolution of matter density perturbations}

\subsection{Superhorizon sized perturbations}
The time variation of dark matter and dark energy densities have
analytic expressions in terms of the expansion factor (Eq.
(\ref{integrate})). In terms of the time variable $\log a$,
analytic solutions can be found for the evolution of superhorizon
sized perturbations. Using this new time variable Eq.
(\ref{ho_cdm}) can be written as
\begin{equation}
\delta_c^{\prime\prime} +
\left(\frac{1-3w_B}{2}+3c^2\right)\delta_{c}^{\prime}+3c^{2}\left(\frac{1-3\omega_B}{2}
\right) \delta_{c} =
\frac{3}{2}(1+3c^2_{s,B})\frac{\delta\rho_B}{\rho_B} \, ,
\label{ho_cdm_a}
\end{equation}
where prime denotes derivatives with respect to $\log a$. The
subindex $B$ stands for background quantities. The behavior of
$\delta\rho_B/\rho_B$ at scales larger than the Jean's length can
be parameterized as:
\begin{equation}
\frac{\delta\rho_B}{\rho_B} =
\left(\frac{\delta\rho_B}{\rho_B}\right)_{H_I}
\left(\frac{a}{a_{H_I}}\right)^{p/2} \, ,
\end{equation}
where $(\delta\rho_B\rho_B)_{H_I}$ is the amplitude of the mode under
consideration at horizon crossing, at time $a_{H_I}$.
After a brief transient period, the evolution of the dark matter and dark energy
perturbations will be given by the inhomogeneous solution associated with  the
time evolution of the gravitational potential:
$\delta_c\sim h \sim a^{p/2}$.
The solutions of eq (\ref{ho_cdm_a}) are:
$p=(4; \; 2-6.6c^2)$ in the
radiation and cold dark matter dominated periods, respectively.
For non-interacting models, the well known solutions are $p=(4; \; 2)$.
These solutions were to be expected; as discussed in \cite{padma},
if $\rho_c\sim a^{-\alpha}$, being $\alpha$ a constant, then
$p\simeq 2(\alpha-2)$. In the radiation epoch, $\alpha =4$, and
in the matter epoch $\alpha=3(1-c^2)$. Thus,
during matter domination, the growth of dark matter density perturbations
slows down with respect to those of non-interacting
models but, in general, the evolution of superhorizon sized
perturbations is not significantly altered by the interaction.

\subsection{Subhorizon sized matter perturbations}
For perturbations inside the horizon, Eqs. (\ref{de_cont}) and
(\ref{de_euler}) have the approximate solution,
\begin{equation}
\delta_x=3\left({\HH\over k}\right)^2c^2r\delta_c \, .
\end{equation}
Since $c^2\ll 1$, we have that
$\Omega_c\delta_c\gg\Omega_x\delta_x$ and the force term in Eq.
(\ref{ho_cdm}) is dominated by the perturbations in the photons
field during the radiation epoch. In the small scale limit
($3\HH/k\ll 1$), $A\simeq 1$ and
\begin{equation}
\ddot \delta_c +\HH (1+3c^2)\dot\delta_c -4\pi
Ga^2\rho_c\left[1+\frac{3c^2r w_{x}}{4\pi G\rho_ca^2(r+1)(1+w_x)}
\right]\delta_c={3\over 2}\HH^2\Omega_\gamma\delta_\gamma .
\label{ho_cdm_simple}
\end{equation}
Even this simplified equation does not have simple analytic
solutions. The slower growth of matter density perturbations in
the IQM compared with non--interacting ones can be understood
analyzing the different coefficients: (A) the interaction
increases the friction term, damping more rapidly the homogeneous
solution and (B) it decreases the gravitational force acting on
the
perturbation. At very early times, when $\Omega_c\le 6c^2%
\mid w_{x}/(1+ w_{x})\mid$, and well within the horizon
perturbations on the photon field oscillate  and
matter perturbations do not grow but undergo damped oscillations \cite{peeb_ratra}.
The characteristic time--scale of the growth of matter density
perturbations is the mean free--fall time, $t_{ff}\sim
(G\rho_c)^{-1/2}$. During the radiation dominated regime the
expansion rate is fixed by Friedmann equation: $H\sim
\sqrt{G\rho_\gamma}$ and since $t_{ff}>>H^{-1}$, perturbations
only grow logarithmically, not as a power law. In our IQM this
effect is more severe. First, at all times the matter density is
smaller than in non--interacting models but with the same values
for the cosmological parameters today. Second, the effective
gravitational force is reduced. Thus, the mean free--fall time
increases and density perturbations grow slower (or get even
erased), compared with a non--interacting model.

Scalar fields coupled to matter would modify gravity inducing an
extra attractive force. A repulsive effect could be obtained by
the exchange of vector bosons. It was first suggested
\cite{amendola-phantom} that phantom scalar fields with
non-standard kinetic term coupled to matter would give rise to a
long-range repulsive force. In our phenomenological model, the
decrease of the gravitational coupling in Eq.~(\ref{ho_cdm_simple}) is
due to our specific ansatz $Q$ for the dark matter --
dark energy interaction.

\subsection{Comparison with other interacting models}
Interacting quintessence models couple dark matter and dark energy
so the energy momentum tensor of the DM and DE are not separately
conserved but obey
$(T^{\mu}_{(\phi)\nu}+T^{\mu}_{(c)\nu})_{;\mu}=0$. In references
\cite{amendola, amendola2004} the coupling is chosen such that
$\dot{\rho_{c}}+3{\cal H} \rho_{c}=(16\pi G/3)^{1/2}\, \beta
\rho_{c}\dot {\phi}_{x}$, where $\phi_{x}$ is the scalar field
describing the dark energy component and $\beta$ the decay rate
coefficient whom, in general, is a time varying function.
By assuming that the scalar field couples to dark
matter only, the evolution of matter density perturbations in the
synchronous gauge is given by
\begin{eqnarray}
\dot{\delta}_{c}&=&-\theta_{c}-{1\over 2}\dot {h} -
\frac{d}{d\tau}(\beta\varphi) \, ,
\label{eq:a1}\\
\dot{\theta}_{c}&=&-\HH(1-2\beta x)\theta_{c}-2\beta k^{2}\phi \,
. \label{eq:a2}
\end{eqnarray}
In this expression, $\varphi=(4\pi G/3)^{1/2}\delta\phi_x$ is the
perturbation in the scalar field and $x$ its kinetic energy. The
evolution of the gravitational field does not depend on the
specific interaction ansatz, and is given by Eq.
(\ref{hgravitation}). For subhorizon sized perturbations,
$\varphi=k^{-2}\HH (\beta\Omega_c\delta_c)$ and, in the radiation
dominated regime, matter perturbations evolve as
\begin{equation}
\ddot\delta _{c}+\HH[1-2\beta x]\dot\delta_{c} -4\pi
Ga^2[1+{4\over 3}\beta^2]\rho_c\delta_c= {3\over 2}\HH^2
2\Omega_\gamma\delta_\gamma \, .\ \label{eq:a3}
\end{equation}
As discussed above, during the radiation period the background
expansion rate is fixed by Friedmann's equation, but the mean
free--fall time is now: $t_{ff}\sim%
[G(1+4\beta^2/3)\rho_c]^{1/2}$. Due to the interaction, the dark
matter density at any given time is smaller than in a
non-interacting model with the same cosmological parameters, and
the difference increases with $\beta$. Likewise, the effective
gravitational constant increases, but since the dependence is
second order in $\beta$, one would expect $t_{ff}$ to be smaller
than in non-interacting models. This statement depends on the
particular interaction ansatz. Since perturbations evolve as if
the  Newton's gravitational constant was a factor $(1+4\beta^2/3)$
larger, the interaction with the scalar field could make density
perturbations to grow faster during the matter dominated regime
due to a larger local gravity. This effect could compensate the
slow growth during the radiation dominated regime and enhancing
the clustering of dark matter perturbations compared with the
uncoupled case, as found in \cite{corasaniti}. But even in this
case, the amplitude of the matter power spectrum was smaller in
the range $(0.01-0.4)h$ Mpc$^{-1}$.

\section{The effect of the Interaction on the Matter Power Spectrum}
In the previous section we have shown that the interaction slows
the growth of matter density perturbations. Only the slower growth
of perturbations in the radiation dominated regime will have a
significant impact on the matter power spectrum today. For
comparison, we shall assume that in interacting and
non-interacting models density perturbations have the same
amplitude when they come within the horizon. For non-interacting
models, this prescription leads to the so called
Harrison--Zeldovich power spectrum \cite{harrison-zeldovich},
characterized by a functional form $P(k)\sim k^{n_s}$ with $n_s=1$
on large scales. During the matter epoch, if density perturbations
evolve with the scale factor as $\delta_{c}\sim a^{p/2}$ and the
background energy density as $\rho_{c}\sim a^{-\alpha}$ (with
$\alpha=$const) the power spectrum will scale with wavenumber as
\begin{equation}\label{A_evol4}
P(k)\sim k^{-3+2p/(\alpha-2)}\, .
\end{equation}
During the matter dominated period $p\simeq 2(\alpha-2)+0.6c^2$ and the slope
of a scale-invariant spectrum is $n_s=1$,
with a very weak dependence on the interaction.
\begin{figure}
\includegraphics[scale=0.9]{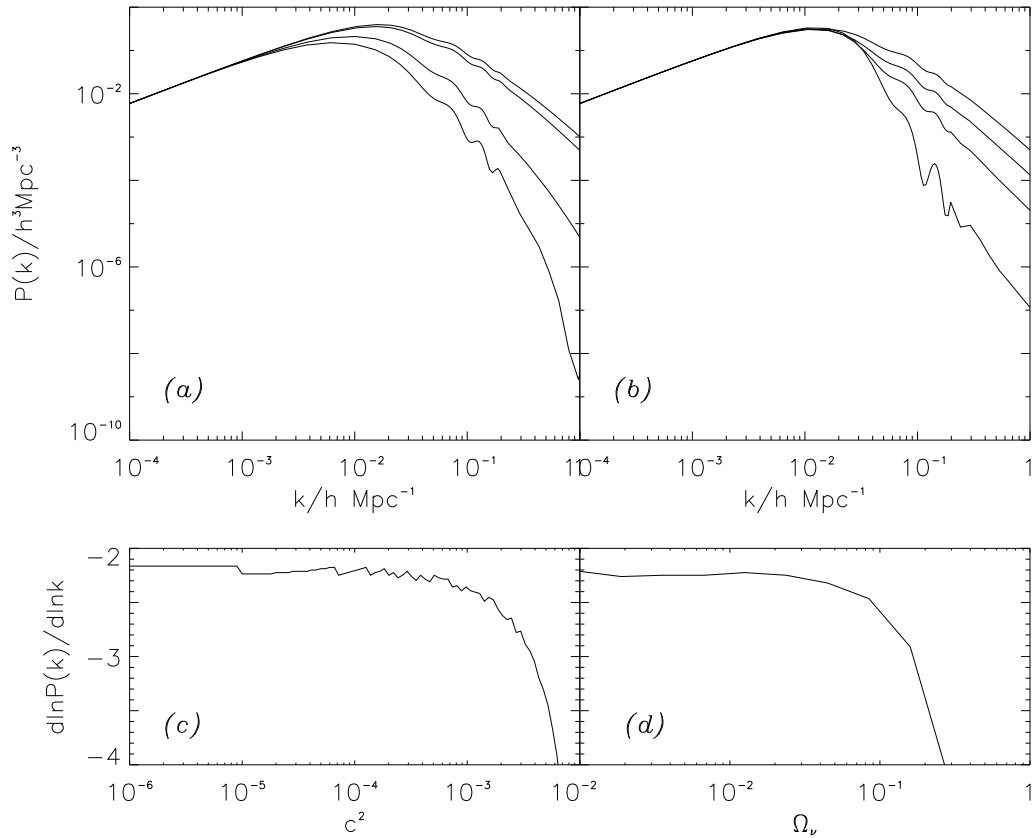}
\vspace*{-1.5cm} \caption{(a) Matter power spectra for the
interacting quintessence model with different rates of energy
transfer. From top to bottom $c^2=0,10^{-3},6\times
10^{-3},10^{-2}$. We took the present value of cosmological
parameters to be: $\Omega_{b}=0.04$, $\Omega_{cdm}=0.23$,
$H_0=72$km s$^{-1}/$Mpc, $\Omega_\Lambda =0.73$ the dark energy
equation of state $w_{x}=-0.9$ and the slope of the matter power
spectrum at large scales $n_s=1$. (b) The same for mixed dark
matter models with one single species of massive neutrinos. From
top to bottom, the fraction of energy density in the form of
neutrinos is: $\Omega_\nu= 0.01, 0.05, 0.1, 0.2$. As before, the
total dark matter energy density was $\Omega_{dm}=0.23$, the rest
was cold dark matter. (c) Variation of the slope of $P(k)$ with
$c^2$, and (d) with massive neutrinos. The slope was computed from
a straight line fit to the data in the interval $k=[0.1,1] h^{-1}$
Mpc.} \label{fig:pk}
\end{figure}

The slope of the matter power spectrum on scales $k\ge k_{eq}$ is
determined by the growth rate of subhorizon sized matter
perturbations during radiation domination.  If a mode that crosses
the horizon before matter--radiation equality ($k\ge k_{eq}$)
grows as $\delta_{c} \sim \tau^{q/2}$ during the radiation
dominated period, then the amplitude of the power spectrum today
would be $P(k)=P(k_{eq})(k_{eq}/k)^{-3+q}$. For cold dark matter
models, dark matter perturbations experience only a logarithmic
growth, so models with less growth will have less power at small
scales as do, for example, mixed dark matter models \cite{davis},
i.e., models containing a significant fraction of massive
neutrinos.

In Fig. \ref{fig:pk}a we plot the power spectrum for different
interacting quintessence models. All models have the cosmological
parameters of the WMAP first year concordance model \cite{wmap}.
From top to bottom, $c^2 = 0,10^{-4},10^{-3},10^{-2}$; the
normalization is arbitrary. Similarly, in Fig. \ref{fig:pk}b we
plot the matter power spectrum of mixed dark matter models with
one species of massive neutrinos, for different neutrino masses:
$m_\nu = 0, 0.1, 1, 10$ eV. With increasing decay rate or neutrino
mass, the matter power spectrum shows larger oscillations, due to
the increased ratio of baryons to dark matter. The slope decreases
with increasing $c^2$ and $m_\nu$. As explained above, potential
wells are shallower with increasing $c^2$; matter perturbations
during radiation domination are damped similarly as do in models
with massive neutrinos. In Fig. \ref{fig:pk}c we plot the change
in the slope of the matter power spectrum as a function of the
energy transfer rate and in \ref{fig:pk}d as a function of the
neutrino mass. As the slope changes smoothly from large to small
scales, for convenience we computed the slope as a straight line
fit to the data in the interval $k=[0.1,1.0]h^{-1}$Mpc. In both
cases the behavior is rather similar: for low values of neutrino
mass and interaction coupling, the slope is approximately $-2$ and
roughly constant. When parameters are increased in either model,
the slope decreases.  Observations of large scale structure that
constrain the neutrino mass can also be used to set constraints on
the strength of DM--DE coupling during the radiation dominated
period. These constraints are complementary to those coming from
skewness of the matter density field, that are sensitive to the
interaction at much lower redshifts \cite{amendola_prl}. Fig.
\ref{fig:pk} shows a significant difference between massive
neutrinos and interacting quintessence: in IQM the maximum of the
matter power spectrum shifts to larger scales. At larger $c^2$,
the dark matter density becomes smaller at any given redshift and
the matter radiation equality is delayed. This does not happen in
models with massive neutrinos where matter-radiation equality
occurs always at the same redshift.

\section{Observational Constraints on Dark Matter - Dark Energy coupling}
Since the interaction affects the slope of the matter power
spectrum, we used the 2dFGRS data \cite{2dgf} to constrain $c^2$.
We used a Monte Carlo Markov chain to run the CMBFAST code,
adapted to solve the IQM described above, through a 7-dimensional
parameter space: ($A$, $\Omega_bh^2$, $\Omega_ch^2$, $H_0$, $n_s$,
$c^2$, $w_{x}$) where $A$ is the normalization of the matter power
spectrum, $\Omega_b$, $\Omega_c$ are the baryon and cold dark
matter fraction in units of the critical density, $n_s$ is the
slope of the matter power spectrum at large scales, $c^2$ measures
the transfer rate of dark energy into dark matter and $w_{x}$ is
the dark energy equation of state parameter. Hereafter $H_0=100h$
km/s Mpc$^{-1}$. It is common practice to call $h$ the Hubble
constant in units of $100$km s$^{-1}/$Mpc and we shall follow this
convention. It should not be confused with the gravitational
potential in the synchronous gauge introduced in Sec. III. We
limit our study to flat models, so the fraction of dark energy is
fixed by the Friedmann equation
$\Omega_x+\Omega_b+\Omega_c+\Omega_\gamma=1$, where
$\Omega_\gamma$ is the photon energy density,  and all densities
are measured in units of the critical density. To simplify, we
studied only adiabatic initial conditions and initial power
spectrum with no running on the spectral index.  We did not
include reionization, or gravitational waves, since they have
little effect on the matter power spectrum.

\begin{figure}
\includegraphics[scale=0.9]{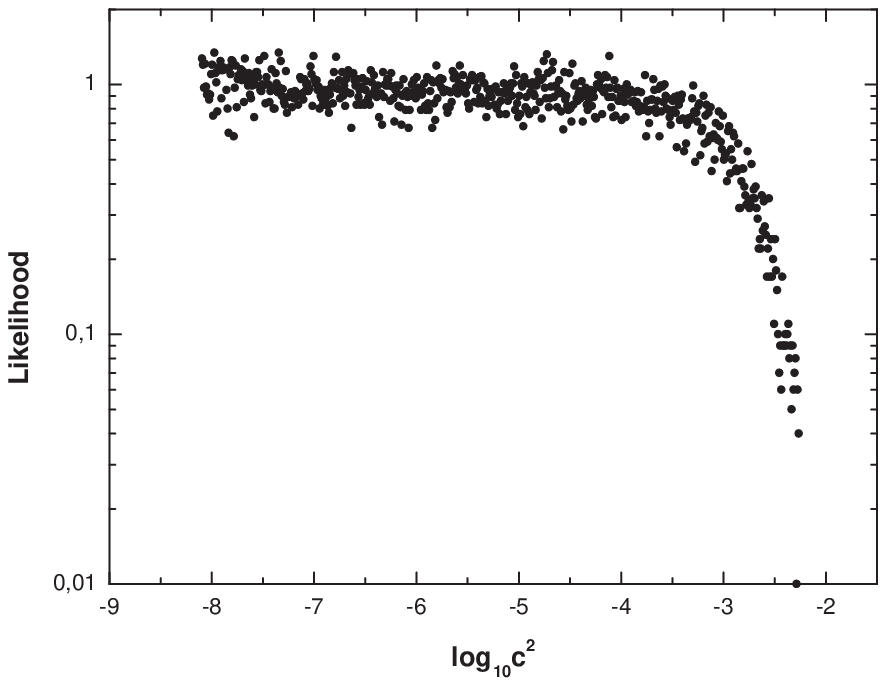}
\vspace*{-1cm} \caption{Marginalized likelihood function for the
2dFGRS data. } \label{fig3}
\end{figure}

Since we are interested in constraining $c^2$ from the shape of
the matter power spectrum, we have to correct for non-linear
effects. We followed the 2dFGRS team and assumed the non-linear
biasing to be well described by
\begin{equation}
P_{gal}(k)=b^2{1+Qk^2\over 1+A_gk}P_{lin,dm}(k) .
\end{equation}
We used  $A_g = 1.4$ and $Q=4.6$. We marginalized over the bias
factor $b$. We did not use the SDSS galaxy power spectrum because
these data were analyzed in real space where non-linear effects
are more important. We used the likelihood codes provided by the
2dFGRS team \cite{2dgf}. As priors, we imposed our chains to take
values within the intervals: $A=[0.5,2.0]$ in units of COBE
normalization, $h=[0.4,1.1]$, $\Omega_bh^2 = [0.00, 0.05]$,
$\Omega_ch^2=[0.0,0.5]$, $n_s=[0.80,1.2]$, $w_x=[-0.5,-1.0]$ and
$c^2=[0,0.05]$. We run the chain for $10^5$ models, that were
sufficient to reach convergence. In Fig.  \ref{fig3} we plot the
marginalized likelihood function obtained from the posterior
distribution of models. The likelihood is very non-gaussian,
reflecting the fact that models do not depend linearly on $c^2$.
The data are rather insensitive to $c^2\le 10^{-3}$ since the
slope does not change significantly up to that value (see Fig.
\ref{fig:pk}c). As discussed above, increasing the interaction
rate leads to smaller fraction of dark matter during the radiation
dominated period and shallower potential wells, larger free--fall
times and, as a result, the amplitude of the matter power spectrum
is damped (see Fig. \ref{fig1}e).

\begin{figure}
{\centering\leavevmode\epsfxsize=\columnwidth\epsfbox{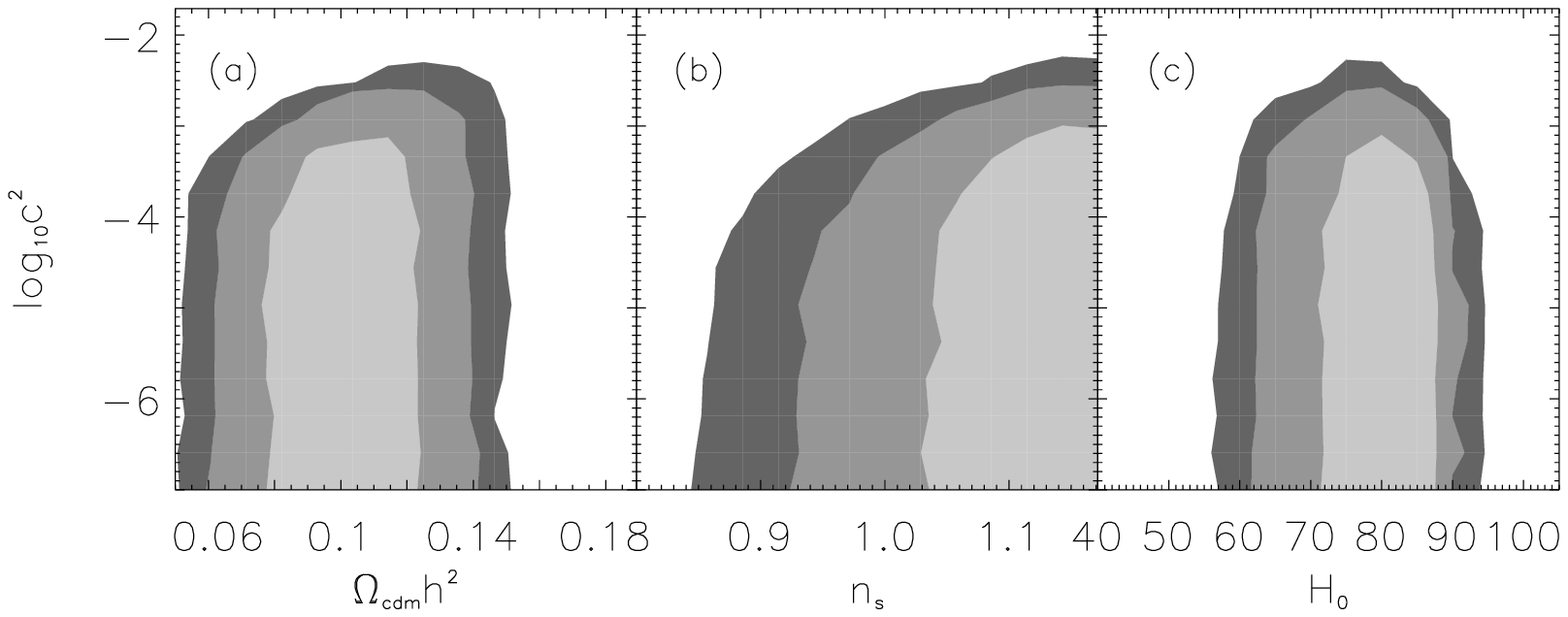}}
\vspace*{-9cm} \caption{Marginalized likelihood function for the
2dFGRS data. } \label{fig4}
\end{figure}

In Fig. \ref{fig4} we show the joint confidence contours at the
68\%, 95\% and 99.99\% level for pairs of parameters after
marginalizing over the rest. Our prior on the spectral index was
too restrictive and did not allow the chains to sample all the
parameter space allowed by the data. Therefore, we can not draw
definitive conclusions about the confidence intervals for all of
the parameters. The figure does show that the data at present do
not have enough statistical power to discriminate the IQM from
non-interacting ones. Our $1\sigma$ confidence levels and upper
limits for the cosmological parameters are $c^2\le
3\times10^{-3}$, $\Omega_c h^2 = 0.1\pm 0.02$, $H_0 =
83^{+6}_{-10}$km s$^{-1}/$Mpc. The data are rather insensitive to
$w_x$ and baryon fraction. Models with $c^2 = 0$ are compatible
with the 2dFGRS data at the $1\sigma$ level. The data show a full
degeneracy with respect to $c^2$ up to $c\simeq 10^{-3}$, in
contrast with the results of \cite{olivares} obtained using the
1st year WMAP data. There the data preferred interacting
quintessence models with respect to non-interacting ones but this
was an artifact of our parameter space since we restricted the
normalization to be that of COBE, penalizing the concordance model
that prefers a lower normalization. A full discussion including
WMAP 3rd year data will be deferred to a forthcoming paper.

\section{Discussion}
Interacting quintessence models have been constructed to solve the
coincidence problem. They make specific predictions that can be
checked against observations of large scale structure. In this
paper we have shown that the interaction induces measurable
effects in the growth of the matter density perturbations and
modifies the power spectrum on small scales.
To summarize, if dark energy decays into dark matter the
background model has less dark matter and more dark energy in the
past compared to non-decaying models. Since the  dark energy does
not cluster on small scales it does not contribute to the growth
of density perturbations. The mean free--fall time increases and
perturbations grow slower than  in non-interacting models. The
slower growth of matter density perturbations during the radiation
and matter period, results on a damping of the matter power
spectrum on those scales that cross the horizon before
matter-radiation equality, but it does not change the slope on
large scales.

The combined effect of shifting the scale of matter radiation equality and changing
the slope of matter power spectrum at small scales is a
distinctive feature of interacting models where the dark energy
does not cluster on small scales. Measurements of matter power
spectrum could eventually reach enough statistical power  to
discriminate between interacting and non-interacting models. For
example, the spectrum obtained from Lymann-$\alpha$ absorption
lines on quasar spectra \cite{lya} probe the matter power spectrum
at redshifts in the interval $(2, 4)$, where non-linear evolution
has not yet erased the primordial information down to megaparsec
scale. The use of more precise information on small and large
scales, could set tighter bounds on the interaction of dark matter
and dark energy.

Our previous results \cite{olivares} and those presented
here indicate that the IQM fits the observational data as well as
non-interacting models, alleviates the coincidence problem and
provides a unified picture of dark matter and dark energy. It
predicts a damping on the matter power spectrum on small scales
that can be used, together with the delay on the matter-radiation
equality, to discriminate it from non-interacting models. The slower
growth of subhorizon sized matter density perturbations within the
horizon provides a clean observational test to proof or rule out a
DM--DE coupling.
 \acknowledgments {This research was partially
supported by the Spanish Ministerio de Educaci\'on y Ciencia under
Grants BFM2003-06033, BFM2000-1322 and PR2005-0359, the ``Junta de
Castilla y Le\'{o}n" (Project SA002/03), and the ``Direcci\'{o}
General de Recerca de Catalunya", under Grant 2005 SGR 00087.}

\end{document}